# Reality-Infused Deep Learning for Angle-resolved Quasi-optical Fourier Surfaces

Wei Chen [1, 2, 3, †], Yuan Gao [1, †], Yiming Yan [1, †], Jiaqing Shen [1, 2], Yongxiang Lin [1], Mingyong Zhuang [1], Zhaogang Dong [2, 3, 4, *], Jinfeng Zhu [1, *]

[1] Institute of Electromagnetics and Acoustics and Key Laboratory of Electromagnetic Wave Science and Detection Technology, Xiamen University, Xiamen, Fujian 361005, China

[2] Institute of Materials Research and Engineering (IMRE), Agency for Science, Technology and Research (A*STAR), 2 Fusionopolis Way, Innovis #08-03, Singapore 138634, Republic of Singapore

[3] Quantum Innovation Centre (Q.InC), Agency for Science Technology and Research (A*STAR), 2 Fusionopolis Way, Innovis #08-03, Singapore 138634, Republic of Singapore

[4] Science, Mathematics, and Technology (SMT), Singapore University of Technology and Design (SUTD), 8 Somapah Road, Singapore 487372

*Corresponding author. E-mail: Zhaogang_dong@sutd.edu.sg; jfzhu@xmu.edu.cn

**Optical Fourier surfaces (OFSs), featuring sinusoidally profiled diffractive elements, manipulate light through patterned nanostructures and incident angle modulation. Compared to altering structural parameters, tuning elevation and azimuth angles offers greater design flexibility for light field control. However, angle-resolved responses of OFSs are often complex due to diverse mode excitations and couplings, complicating the alignment between simulations and practical fabrication. Here, we present a reality-infused deep learning framework, empowered by angle-resolved measurements, to enable real-time and accurate predictions of angular dispersion in quasi-OFSs. This approach captures critical features, including nanofabrication and measurement imperfections, which conventional simulation-based methods typically overlook. Our framework significantly accelerates the design process while achieving predictive performance highly consistent with experimental observations across broad angular and spectral ranges. Our study supports valuable insights into the development of OFS-based devices, and represents a paradigm shift from simulation-**

driven to reality-infused methods, paving the way for advancements in optical design applications.

**Teaser:** A reality-infused deep learning paradigm that leverages experimental data to inherently capture real-world imperfections and variability in optical metasurfaces.

**Introduction**

Optical Fourier surfaces (OFSs), characterized by sinusoidally profiled diffractive optical elements, are carefully engineered to manipulate light with high efficiency. Their growing importance is driven by recent advances in nanofabrication technologies and the increasing demand for novel optical functionalities *(1-3)*. Despite their theoretical appeal, the fabrication of ideal OFSs with perfect sinusoidal profiles demands highly sophisticated and costly nanofabrication techniques, which are currently impractical for wafer-scale production. In most experimental settings— particularly those involving scalable fabrication methods such as nanoimprint lithography—the resulting surface geometries deviate from ideal sinusoidal forms and instead exhibit intermediate profiles between binary and sinusoidal structures. We refer to this class of practically attainable structures as quasi-optical Fourier surfaces (quasi-OFSs), as shown in Fig. 1A *(4)*. Quasi-OFSs approximate the functional characteristics of ideal OFSs while offering compatibility with large-area, high-throughput manufacturing. The grating vector of a patterned OFS ( $G_m = 2\pi m/\Lambda$ ) represents a specific spatial frequency, where $\Lambda$ is period and $m$ denotes order *(5)*. These deviations introduce discrepancies between the actual and expected optical responses, which become particularly critical when angular momentum space is considered *(6)*. Under angle-resolved conditions, incident light with wavevector $k_i$ interacting with a periodic nanostructure undergoes diffraction governed by the condition $k_i \pm mG_m = k_{s,m}$, where $k_i$, $G_m$, $k_{s,m}$ and $m$ represent the light vector, grating vector, *k*-vector of constructively scattered light, and diffraction order, respectively, as shown in **Fig. 1A** *(4-5)*. Therefore, the inclusion of angular degrees of freedom introduces additional mode excitations and inter-mode coupling, leading to increasingly intricate light–matter interactions. This complexity presents major challenges in simultaneously coordinating accurate electromagnetic simulation and reliable nanofabrication. On the one hand, conventional simulation-based design approaches often suffer from numerical instability or divergence under oblique incidence, particularly at large angles where the computation becomes prohibitively intensive *(7)*. On the other hand, the introduction of angular variables of sources alongside geometric parameters dramatically amplifies the degree of freedom and complexity of the OFS and quasi-OFS development.

Recent breakthroughs in artificial intelligence have brought deep learning (DL) to the forefront of optical design, offering fast and flexible data-driven alternatives to traditional computational methods [8–24]. Most existing DL strategies rely on electromagnetic simulations to generate large datasets for model training. However, conventional DL (C-DL) architectures are inherently limited by their dependence on simulated data and are unable to capture real-world imperfections such as fabrication defects and measurement inconsistencies. As a result, C-DL predictions often diverge significantly from experimental observations—particularly for large-angle incidence, which plays a critical role in defining quasi-OFS optical behavior [25–26]. Moreover, the acquisition of large-scale experimental datasets is often expensive, time-consuming, and technically challenging [19], posing a major bottleneck for data-driven design paradigms.

To bridge the persistent gap between simulation and experiment, we propose a reality-infused DL (R-DL) design paradigm that leverages experimental data, rather than simulation results. This strategy enables the model to inherently capture real-world imperfections and variability *(27)*. As demonstrated in Fig. 1B, our framework significantly improves prediction accuracy while retaining ultrafast computational efficiency compared to existing design paradigms (see Table S1). The proposed architecture, illustrated in Fig. 1C, integrates both inverse and forward design processes and employs a series of OFS-like metagratings as proof-of-concept examples. We utilize a set of high-throughput, Fourier-optics-based angle-resolved imaging spectroscopy (ARS) to rapidly generate both general and plentiful datasets, which can be widely applied in the field of nanophotonics. In our paradigm, incident angle and azimuthal orientation are incorporated as tunable design variables, effectively transforming the design task from structure-only optimization to joint structure-angle co-design. The learning capabilities of the R-DL architecture are systematically investigated using the experiment-based transformer model (ExpForm). Compared to simulation methods, our framework significantly improves consistency with experimental results and accelerates the optical design process of metasurfaces. This strategy presents a flexible and powerful approach to eliminate the mismatch between the design and fabrication, facilitating further advancements in angle-resolved meta-devices.

To bridge the persistent gap between simulation and experiment, we propose a reality-infused DL (R-DL) design paradigm that leverages experimental data, rather than simulation results. This strategy enables the model to inherently capture real-world imperfections and variability *(27)*. As demonstrated in Fig. 1B, our framework significantly improves prediction accuracy while retaining ultrafast computational efficiency compared to existing design paradigms (see Table S1). The proposed architecture, illustrated in Fig. 1C, integrates both inverse and forward design processes and employs a series of OFS-like metagratings as proof-of-concept examples. We utilize a set of high-throughput, Fourier-optics-based angle-resolved imaging spectroscopy (ARS) to rapidly generate both general and plentiful datasets, which can be widely applied in the field of nanophotonics. In our paradigm, incident angle and azimuthal orientation are incorporated as tunable design variables, effectively transforming the design task from structure-only optimization to joint structure-angle co-design. The learning capabilities of the R-DL architecture are systematically investigated using the experiment-based transformer model (ExpForm). Compared to simulation methods, our framework significantly improves consistency with experimental results and accelerates the optical design process of metasurfaces. This strategy presents a flexible and powerful approach to eliminate the mismatch between the design and fabrication, facilitating further advancements in angle-resolved meta-devices.

## Results

### Fabrication, measurement and theoretical analysis of metasurfaces

Before exploring the intelligent design facilitated by the R-DL strategy, it is essential to establish our proof-of-concept model *(28-29)*. We fabricate four typical nanopatterns on the silicon wafers using nanoimprint lithography and deposit a 150 nm-thick silver layer by magnetron sputtering (see Fig. S1 and Table S1). We measure reflectance spectra via the ARS system shown in **Fig. 2A,** where BS, L1/L2, M, and LS denote the beam splitter, lenses, mirror, and light source *(30-31)* (see more details in Method). The dispersions from the visible to near-infrared ranges are observed through the two-dimensional charged-coupled-device camera *(32)*. The photograph of four quasi-OFS samples with each area of 2.5 cm × 2.5 cm on a wafer exhibits distinct visual effects of diffraction. The measured reflectance spectra for $\theta$=0° and $\varphi$=0°, along with corresponding images of atomic force microscopy (AFM) profile and scanning electron microscopy (SEM), are provided in Fig. 2B. The reflectance spectra display multiple resonance peaks due to the excitation of various modes and optical couplings. Notably, the significant resonance valleys have a gradual redshift as the period becomes smaller, deviating from surface plasmonic resonances *(33)*. To clarify the resonance mechanism, we perform finite-difference time-domain (FDTD) modelling and simulate based on the morphological characterization and measurement (see Fig. S2-S3), as shown in Fig. 2C. To distinguish between resonance and diffraction modes, we calculate electric field distributions in the *x-z* plane at resonance wavelengths. For modes of P1 and P3, the field enhancement of more than an order of magnitude is confined within the grating grooves (cavities), representing cavity modes (CMs) *(34-35)*. For the P2 mode, the field is diffracted from the grating surface, corresponding to Rayleigh's anomaly (RA) *(36)*. When a diffraction order ($m$, $n$) appears or disappears at the grazing angle, the RA occurs, causing abrupt changes in the reflectance spectra. For a structure with period $\Lambda$, RA can be predicted by the dispersion relation *(37)*,

$$\left(m\frac{\lambda}{\Lambda}\right)^2 + \left(n\frac{\lambda}{\Lambda}\right)^2 + 2\sin\theta\left(m\frac{\lambda}{\Lambda}\cos\varphi + n\frac{\lambda}{\Lambda}\sin\varphi\right) - \cos^2\theta = 0 \tag{1}$$

As shown in Eq. (1), the RA mainly depends on the grating period $\Lambda$ at normal incidence, and the expression can be simplified as below:

$$\left(m\frac{\lambda}{\Lambda}\right)^2 + \left(n\frac{\lambda}{\Lambda}\right)^2 = 1 \tag{2}$$

Therefore, the (1, 0) order of RA is assumed to act at the wavelength of 820 nm, which agrees with the P2 wavelength of 848.8 nm. In fact, here the grating is thick enough to support multiple CMs in the given wavelength range *(38)*. In this case, the resonance wavelength $\lambda = 2h \cdot n_{neff}/m$, where $m$ stands for the CM order, $n_{neff}$ is the effective refractive index (ERI) of the excited TM mode inside the cavity (see Fig. S4) *(36)*. Furthermore, the disappearance of resonances when the modes converge suggests the formation of Friedrich-Wintgen bound states in the continuum (BIC). A key feature of the Friedrich-Wintgen BIC in the proposed structure is its origin from the interference between two distinct resonances *(39)*. Without strong coupling, their dispersions simply intersect at a specific angular position. The quality factor of this quasi-BIC mode shown on the spectrum is not evidently high, which is constrained by the intrinsic optical losses of the materials *(40-42)*. According to the above brief analyses, introducing the angular variables $\theta$ and $\varphi$, in addition to the structural parameters ($w_1$, $w_2$, $p$, $h$), would significantly enhance the design complexity.

**Reality-infused DL neural networks for metasurface design**

On this basis, we establish the R-DL architecture based on an experimental transformer model, termed ExpForm. As depicted in **Figure 3**, the framework consists of the forward neural network

(FNN) and inverse neural network (INN). In the FNN, the input is the OFS model vector $S$ $[\lambda_1,$ $\lambda_2..., \lambda_{2636}]$ from 380 to 1100 nm and the output is the ARS-measured spectral vector $S$. The role of the FNN is to quickly predict the optical spectrum, effectively replacing time-consuming simulations. In the FNN, the input data consists of the parameter vector $M$, which undergoes input embedding and positional encoding before being passed into a transformer encoder composed of $L$ stacked layers. The final prediction of the spectral response $S$ is obtained after processing through a fully connected layer. In the INN, the input corresponds to a spectral vector $S$ for 2636 wavelength points ranging from 380 to 1100 nm, while the output represents the parameter vector $M$. To match input and output dimensions, we segment $S$ into $N$ smaller patches. We perform a convolutional embedding on each patch, converting spectral segments into a set of feature vectors. Then we incorporate positional embeddings into these vectors. The sequence of embedded vectors is then fed into a transformer encoder. The ExpForm encoder consists of $L$ identical layers, each containing a multi-head self-attention mechanism (with $H$ heads) and a position-wise feed-forward network, along with residual connections and layer normalization. We adopt the following equation to describe the self-attention mechanism *(20)*,

$$\text{Attention(Q, K, V)} = \text{softmax}\left(\frac{QK^T}{\sqrt{d_k}}\right)V \tag{3}$$

The self-attention mechanism operates through three learnable matrices: query ($Q$), key ($K$), and value ($V$), derived from linear transformations of the input sequence, where $d_k$ is the dimension of these matrices. The dot products between $Q$ and all keys ($K$) are divided by $\sqrt{d_k}$, and the weights on the values can be obtained via softmax function. The multi-head attention mechanism generates different representations of ($Q$, $K$, $V$), which can produce diverse attention patterns for the heads on each layer. Each head attention can be described as follows *(20)*,

$$\text{head}_i = \text{Attention}\left(QW_i^Q, KW_i^K, VW_i^V\right) \tag{4}$$

$$\text{MultiHead}(Q, K, V) = \text{Concat}\left(\text{head}_1, ... \text{head}_i, ... \text{head}_M\right)W^O \tag{5}$$

Here the $W_i^Q$, $W_i^K$, $W_i^V$ are weight parameter matrices for $Q$, $K$, and $V$, respectively. There are $H$ heads for the attention mechanism. The $head_i$ represents the $i^{th}$ head attention. $W^O$ denotes the weight parameter matrix related to all the heads. By calculating each head attention and concatenating the results, we can obtain the results of the multi-head attention mechanism. The fully connected feed-forward network, which is applied to each position identically. We train the neural network by minimizing the mean square error (MSE) between the predictions and ground truth, described as

$$MSE = \frac{1}{N_s} \sum_{i=1}^{L} \left( T_i - \tilde{T}_i \right)^2 \tag{6}$$

where $T_i$ denotes the spectra or model vectors predicted by the NNs, $\tilde{T}_i$ is the measured spectra or parameters, and $N_s$ is the number of spectral sampling points. We randomly pick up an instance with $M$ [$\varphi$=72.5, $\theta$=21.9, $p$=1050, $h$=572, $w_1$=610, $w_2$=900] from testing sample space and compare the simulation results with experimental measurements in Fig. 3B. For this instance with oblique incident and azimuthal angles, the simulation fails to capture matter-of-fact spectral features, including the wavelength, shape, and quality factor of optical resonance. This disparity between simulation and experiment can be attributed to a variety of factors, such as surface and grain boundary scattering, machining tolerances, and other fabrication imperfections *(44-51)*. As shown in Fig. 3C, our ExpForm model significantly reduces absolute spectral error across most wavelength ranges, particularly in the near-infrared region. More importantly, compared to conventional simulations, our method achieves a remarkable 900-fold increase in design efficiency (number/hour) while improving measurement agreement to 99.79%, as highlighted in Fig. 3D. To emphasize the groundbreaking nature of our proposed R-DL framework, Fig. 3E presents a comparative analysis of simulation and experimental results against state-of-the-art studies. Traditional methods like parameter scanning, algorithm optimization, and simulation-driven DL often struggle to achieve a high level of consistency with experimental results typically. Specifically, in reference *(26)*, DL predictions exhibit an accuracy of <30% compared to experimental results. Moreover, it is widely recognized that wide-angle simulations drastically increase computational time. Our method can rapidly obtain measured spectra at different incident angles, benefiting from high-throughput Fourier-optics-based angle-resolved imaging technology.

**Demonstration of angle-resolved prediction for metasurfaces**

To further validate our approach, we employ the ExpForm model to predict the angle-resolved energy bands of metasurface samples in real time. This capability, previously unattainable with conventional simulation-based methods, is demonstrated as a proof of concept by analyzing samples II and IV (corresponding to **Fig. 4A** and Fig. 4B) at different azimuthal angles. Depending on the orientation of the electric field vector relative to the incident light, the quasi-OFSs are either vertically or horizontally aligned. Firstly, we randomly selected two instances to compare the predicted and experimental spectra at specific angles. It is clear that ExpForm accurately predicts each resonant mode. We further compare the angle-resolved spectra predicted by ExpForm with the measured results. The ExpForm framework shows remarkable consistency with these experimental measurements (Fig. 4C and Fig. 4D). For a more intuitive comparison of the accuracy, we display the errors with respect to the ground truth. Our ExpForm predictions maintain a slight error across a wide range of wavelengths and angles. This precision will substantially accelerate research into the optical diffraction properties of meta-devices.

**Reality-infused DL neural networks for on-demand design**

To highlight the advanced capabilities of our model in on-demand spectral customization, we further investigate the learning performance of ExpForm in the INN framework, where the spectra $S$ [$\lambda_1, \lambda_2..., \lambda_{2636}$] are split into $N$ patches. The inverse network first predicts the structural and angular parameters that meet the target spectral requirements, and then the forward network verifies the accuracy by predicting the corresponding spectrum, which users can compare against their desired response. **Figure 5** presents multiple inverse design examples generated via our ExpForm model. As shown in Fig. 5A to Fig. 5B, we successfully designed and experimentally realized single narrowband resonances at the desired wavelengths of 700 nm and 1000 nm. Compared to conventional inverse design approaches that rely on simulated structural parameter, our angle-resolved method offers a more flexible alternative. By simply tuning the incident and azimuthal angles, it enables the realization of a wide range of target spectra without the need to redesign the physical structure. Moreover, Fig. 5C demonstrates our model's ability to flexibly switch between low-reflection and high-reflection spectral profiles, with the re-predicted spectra maintaining excellent agreement with the ground truth, confirming the robustness of our approach. In Fig. 5D, we further showcase the inverse design of dual-band resonances at two selected wavelengths, demonstrating the model's capability to handle more complex spectral requirements. These representative cases illustrate the predicted structures and their experimentally realized spectra for single-, dual-resonance, and high-reflection targets. The strong agreement between the target spectra and measured spectra validates the effectiveness of our inverse design process. To further enhance accessibility, we have developed a user-friendly software interface for real-time inverse design, enabling users to interactively specify desired spectral features. This tool significantly simplifies the design process, bridging the gap between target optical responses and fabrication parameters. A screenshot of the interface is provided in Fig. S6-S8, and a video demonstration showcasing its functionality has been uploaded as Supplementary Multimedia. This interactive platform underscores the practical impact of our approach, making real-time, on-demand metasurface design more accessible to the broader scientific community.

**Discussion**

In summary, we introduce a reality-infused deep learning framework that leverages high-throughput angle-resolved imaging spectroscopy to invigorate the development of metasurface-based devices via changing incident and azimuthal angles. Our method can rapidly capture measured spectra with plentiful angular dispersion features according to experiment results, a capability not achievable with traditional simulation-based methods. Systematical evaluations demonstrate that the reality-infused deep learning framework effectively perceives the experimental parameter space and executes high-precision forward and inverse predictions. Compared with the conventional method, our architecture not only significantly accelerates design efficiency but also dramatically improves the experimental agreement. This strategy offers a practical and efficient approach to integrate experimental data into the design workflow, providing valuable guidance for the development of metasurface and Fourier surface. Furthermore, it establishes a transformative paradigm for deep learning applications in optical devices design and beyond.

## Materials and Methods

**Numerical simulations.** The numerical simulation is performed using the Finite-Difference Time Domain (FDTD) Solutions, Lumerical software. The optical parameters of silver are obtained from the Handbook of Chemistry and Physics *(52)*. Based on SEM and AFM analyses, we incorporated a certain degree of fabrication-induced disorder and roughness into our simulations. For normal incidence, the boundary conditions along the $x$-axis and $y$-axis are periodic, while the z-axis has perfectly matched layers. For oblique incidence, the boundary conditions along the $x$-axis and $y$-axis are based on the broadband fixed angle source technique, with the $z$-axis having perfectly matched layers. Plane waves propagate along the $z$-axis to the metasuface and the normalized reflectance is collected by a frequency-domain monitor of field and power placed behind the excitation source.

**Optical measurements.** Reflectance spectra of the fabricated samples are experimentally characterized to investigate their angular dispersive optical responses. Collimated white light from a 100 W halogen source is incident on the sample surface, and the angle- and polarized-resolved reflectance spectra are measured with a Fourier-optics-based angle-resolved imaging spectrometer (ARMS, Ideaoptics, China). The sample is illuminated at the front focal plane of the microscope by a halogen lamp using Köhler illumination. The incident beam, convergent and linearly polarized, passes through the incident linear polarizer and is focused by an objective lens. Subsequently, the reflected beam is Fourier transformed by the same objective lens and imaged by an imaging spectrometer. Reflectance spectra are normalized to a mirror reference over the working wavelengths. Due to the presence of noise in the measured signal, we employ the Savitzky-Golay method to smooth the curves, and the comparison of curves before and after smoothing is depicted in Fig. S9. The morphological evaluation of the samples is conducted using the AFM (Dimension Icon, Bruker, Germany) and the SEM (SIGMA HD, Carl Zeiss, Germany).

**Sample fabrication.** The samples are fabricated by nanoimprint lithography, which can be referred to the previous work *(53-54)*. First, nanoimprint lithography stamps periodic nanopatterns onto a UV-curable resist on a silicon substrate. Then, silicon etching creates silicon nanopatterns. Finally, the residual resist is removed with 60 s of oxygen plasma treatment. Subsequently, a 150 nm-thick silver layer is deposited on the silicon wafer by magnetron sputtering (DISC-SP-3200, Chuangshiweina Technology, Beijing, China) for 150s (power: 400 W; argon flow rate: 60 sccm), respectively.

**Deep learning.** We acquire the high-throughput training dataset using the ARS system, which captures angle-dependent dispersion spectra with each measurement taking approximately 6 minutes. We collect 25232 spectral instances from measurement experiments, and 90% are blindly selected as the training samples, the remaining 10% are used for validation and testing samples. The sample ranges can be found in Table S1. The model is constructed based on the open-source machine-learning framework of PyTorch. The version we used is Python3.8 and the desktop is a Windows 10 operation system based on the GeForce GTX 3080 GPU, Intel(R) Core(TM) i7-

10700K CPU @ 3.80GHz 3.79 GHz, and 12GB of RAM. We adopt the transformer model and multilayer perceptron model to build our proposed reality-infused scheme.

**Data Availability Statement**

The full pre-training experimental dataset and all angle-resolved spectral measurements used in this study have been made publicly available. This open-access resource is intended to help other research groups reduce the substantial workload involved in constructing large datasets and to enable fair comparisons of different neural network models on a shared training set. The data can be accessed on Zenodo (https://doi.org/10.5281/zenodo.15073297).

**Supplementary Materials**

Fig. S1. Flow schemes of (a) the thermal-NIL and (b) the UV-NIL process. (c) SEM images of four imprinting templates used in this study.

Fig. S2. (a) Schematic of the sample wafer. (b) Photograph of the sample wafer. (c) Cross-sectional Scanning Electron Microscope (SEM) image for Sample IV, showcasing the quasi-trapezoidal structure of the nanogratings.

Fig. S3. Atomic Force Microscope images of our fabricated samples.

Fig. S4. Reflectance spectra with different periods.

Fig. S5. 3D $P_{abs}$ distribution for degree-of-disorder $\sigma$=0. For the simulation of 650 nm 2D $P_{abs}$, the x-y plane monitor is placed on the top aluminium film surface, while for the 1550 nm 2D $P_{abs}$, the x-y plane monitor is placed on the bottom aluminium film surface.

Fig. S6. Scanning electron microscope image of the disordered metasurfaces with UCNPs.

Fig. S7. The experimental setup used for the luminescence characterization.

Fig. S8. Upconversion emission as a function of different testing areas for the disordered metasurfaces without UCNP.

Fig. S9. Dark current without laser light in the reverse bias condition.

Supplementary Text. Sample preparation process, range of training parameters and performance comparison of different design methods

**References**
[1] Holsteen A L, Cihan A F, Brongersma M L. Temporal color mixing and dynamic beam shaping with silicon metasurfaces. Science, 2019, 365(6450): 257-260.
[2] Ardizzone V, Riminucci F, Zanotti S, et al. Polariton Bose-Einstein condensate from a bound state in the continuum. Nature, 2022, 605(7910): 447-452.
[3] Chen Y, Deng H, Sha X, et al. Observation of intrinsic chiral bound states in the continuum. Nature, 2023, 613(7944): 474-478.
[4] Lassaline N, Brechbühler R, Vonk S J W, et al. Optical fourier surfaces. Nature, 2020, 582(7813): 506-510.
[5] Lim Y, Hong S J, Cho Y D, et al. Fourier Surfaces Reaching Full - Color Diffraction Limits. Advanced Materials, 2024, 36(40): 2404540.
[6] Lim Y, Park H, Kang B, et al. Holography, Fourier optics, and beyond photonic crystals: Holographic fabrications for Weyl points, bound states in the continuum, and exceptional points. Advanced Photonics Research, 2021, 2(8): 2100061.
[7] Liu, Yicheng, et al. "TO-FDTD method for arbitrary skewed periodic structures at oblique incidence." IEEE Transactions on Microwave Theory and Techniques 68.2 (2019): 564-572.
[8] Xiong B, Liu Y, Xu Y, et al. Breaking the limitation of polarization multiplexing in optical metasurfaces with engineered noise. Science, 2023, 379(6629): 294-299.
[9] Cai, Guiyi, et al. "Compact angle-resolved metasurface spectrometer." Nature Materials 23.1 (2024): 71-78.
[10] Dong Z, Jin L, Rezaei S D, et al. Schrödinger's red pixel by quasi-bound-states-in-the-continuum. Science Advances, 2022, 8(8): eabm4512.
[11] Jiang, J; Fan, J A. Global optimization of dielectric metasurfaces using a physics-driven neural network. Nano Letters, 19, 2019, 5366-5372.
[12] Peurifoy J, Shen Y, Jing L, et al. Nanophotonic particle simulation and inverse design using artificial neural networks. Science Advances, 2018, 4(6): eaar4206.
[13] Lin X, Rivenson Y, Yardimci N T, et al. All-optical machine learning using diffractive deep neural networks. Science, 2018, 361(6406): 1004-1008.
[14] Ma W, Chen W, Li D, et al. Deep learning empowering design for selective solar absorber. Nanophotonics, 2023, 12(18): 3589-3601.
[15] Qian C, Zheng B, Shen Y, et al. Deep-learning-enabled self-adaptive microwave cloak without human intervention. Nature Photonics, 2020, 14(6): 383-390.
[16] Ruan, Qifeng, et al. "Growth of monodisperse gold nanospheres with diameters from 20 nm to 220 nm and their core/satellite nanostructures." Advanced Optical Materials 2.1 (2014): 65-73.
[17] Zhu R, Qiu T, Wang J, et al. Phase-to-pattern inverse design paradigm for fast realization of functional metasurfaces via transfer learning. Nature Communications, 2021, 12(1): 2974.
[18] Ding Z, Su W, Luo Y, et al. Artificial neural network-based inverse design of metasurface absorber with tunable absorption window. Materials & Design, 2023, 234: 112331.
[19] Piccinotti, Davide, et al. "Artificial intelligence for photonics and photonic materials." Reports on Progress in Physics 84.1 (2020): 012401.
[20] Chen W, Li Y, Liu Y, et al. All-Dielectric SERS Metasurface with Strong Coupling Quasi-BIC Energized by Transformer-Based Deep Learning. Advanced Optical Materials, 2024, 12(4): 2301697.
[21] Ma W, Liu Z, Kudyshev Z A, et al. Deep learning for the design of photonic structures. Nature Photonics, 2021, 15(2): 77-90.

[22] Xiong J, Shen J, Gao Y, et al. Real-Time On-Demand Design of Circuit-Analog Plasmonic Stack Metamaterials by Divide-and-Conquer Deep Learning. Laser & Photonics Reviews, 2023, 17(3): 2100738.

[23] Wiecha P R, Muskens O L. Deep learning meets nanophotonics: a generalized accurate predictor for near fields and far fields of arbitrary 3D nanostructures. Nano Letters, 2019, 20(1): 329-338.

[24] Wang L, Ruan Z, Wang H, et al. Deep learning based recognition of different mode bases in ring-core fiber. Laser & Photonics Reviews, 2020, 14(11): 2000249.

[25] Chen W, Gao Y, Li Y, et al. Broadband Solar Metamaterial Absorbers Empowered by Transformer-Based Deep Learning. Advanced Science, 2023, 10(13): 2206718.

[26] Ma X, Ma Y, Cunha P, et al. Strategical deep learning for photonic bound states in the continuum. Laser & Photonics Reviews, 2022, 16(10): 2100658.

[27] Garribba, Lorenza, et al. "Short-term molecular consequences of chromosome mis-segregation for genome stability." Nature Communications 14.1 (2023): 1353.

[28] Chan, John You En, et al. "Full geometric control of hidden color information in diffraction gratings under angled white light illumination." Nano Letters 22.20 (2022): 8189-8195.

[29] Wang H, Wang H, Ruan Q, et al. Coloured vortex beams with incoherent white light illumination. Nature Nanotechnology, 2023, 18(3): 264-272.

[30] Wang B, Liu W, Zhao M, et al. Generating optical vortex beams by momentum-space polarization vortices centred at bound states in the continuum. Nature Photonics, 2020, 14(10): 623-628.

[31] Lassaline N, Brechbühler R, Vonk S J W, et al. Optical fourier surfaces. Nature, 2020, 582(7813): 506-510.

[32] Li T, Chen A, Fan L, et al. Photonic-dispersion neural networks for inverse scattering problems. Light: Science & Applications, 2021, 10(1): 154.

[33] Liang H, Wang X, Li F, et al. Label-free plasmonic metasensing of PSA and exosomes in serum for rapid high-sensitivity diagnosis of early prostate cancer. Biosensors and Bioelectronics, 2023, 235: 115380.

[34] Kazemi-Zanjani N, Shayegannia M, Prinja R, et al. Multiwavelength Surface-Enhanced Raman Spectroscopy Using Rainbow Trapping in Width-Graded Plasmonic Gratings. Advanced Optical Materials, 2018, 6(4): 1701136.

[35] Genevet P, Tetienne J P, Gatzogiannis E, et al. Large enhancement of nonlinear optical phenomena by plasmonic nanocavity gratings. Nano Letters, 2010, 10(12): 4880-4883.

[36] Abutoama M, Bajaj A, Li D, et al. Resonant modes of reflecting gratings engineered for multimodal sensing. APL Photonics, 2020, 5(7): 076108.

[37] Maystre D. Theory of Wood's anomalies//Plasmonics: from basics to advanced topics. Berlin, Heidelberg: Springer Berlin Heidelberg, 2012: 39-83.

[38] Quaranta G, Basset G, Martin O J F, et al. Recent advances in resonant waveguide gratings. Laser & Photonics Reviews, 2018, 12(9): 1800017.

[39] Azzam S I, Shalaev V M, Boltasseva A, et al. Formation of bound states in the continuum in hybrid plasmonic-photonic systems. Physical Review Letters, 2018, 121(25): 253901.

[40] Khattou S, Rezzouk Y, Amrani M, et al. Friedrich-Wintgen bound states in the continuum in a photonic and plasmonic T-shaped cavity: Application to filtering and sensing. Physical Review Applied, 2023, 20(4): 044015.

[41] Hentschel M, Koshelev K, Sterl F, et al. Dielectric Mie voids: confining light in air. Light: Science & Applications, 2023, 12(1): 3.

[42] Zheng M, Shen Y, Zou Q, et al. Moisture-Driven Switching of Plasmonic Bound States in the Continuum in the Visible Region. Advanced Functional Materials, 2023, 33(3): 2209368.

[43] Chen Y, Zhu J, Xie Y, et al. Smart inverse design of graphene-based photonic metamaterials by an adaptive artificial neural network. Nanoscale, 2019, 11(19): 9749-9755.

[44] Shen Y, He K, Zou Q, et al. Ultrasmooth gold nanogroove arrays: Ultranarrow plasmon resonators with linewidth down to 2 nm and their applications in refractive index sensing. Advanced Functional Materials, 2022, 32(10): 2108741.

[45] Malkiel I, Mrejen M, Nagler A, et al. Plasmonic nanostructure design and characterization via deep learning. Light: Science & Applications, 2018, 7(1): 60.

[46] Ho J, Dong Z, Leong H S, et al. Miniaturizing color-sensitive photodetectors via hybrid nanoantennas toward submicrometer dimensions. Science Advances, 2022, 8(47): eadd3868.

[47] Wang B, An R, Chan E A, et al. Retrieving positions of closely packed subwavelength nanoparticles from their diffraction patterns. Applied Physics Letters, 2024, 124(15).

[48] Jiang L, Yin T, Dubrovkin A M, et al. In-plane coherent control of plasmon resonances for plasmonic switching and encoding. Light: Science & Applications, 2019, 8(1): 21.

[49] Xu J, Dong Z, Asbahi M, et al. Multiphoton upconversion enhanced by deep subwavelength near-field confinement. Nano Letters, 2021, 21(7): 3044-3051.

[50] Abdelraouf O A M, Anthur A P, Dong Z, et al. Multistate tuning of third harmonic generation in fano-resonant hybrid dielectric metasurfaces. Advanced Functional Materials, 2021, 31(48): 2104627.

[51] Lim Y, Park H, Kang B, et al. Holography, Fourier optics, and beyond photonic crystals: Holographic fabrications for Weyl points, bound states in the continuum, and exceptional points. Advanced Photonics Research, 2021, 2(8): 2100061.

[52] Haynes, William M. CRC handbook of chemistry and physics. CRC press, 2016.

[53] Hu, Rongsheng, et al. "Exploring aptamer-based metasurfaces for label-free plasmonic biosensing of breast tumor-derived exosomes." Advanced Optical Materials 12.28 (2024): 2401180.

[54] Li, Fajun, et al. "Plasmonic Nanograin Metasurface with Disorder-Enhanced Biosensing for SARS-CoV-2 Variant and Antibodies." Advanced Functional Materials (2024): 2401983.

## Acknowledgments

**Funding:**

The authors thank Prof. L. Shi and Jian Zi from Fudan University, and Dr. Haiwei Yin from Ideaoptics Inc. for the support on the angle-resolved spectroscopy measurements. This work was supported by the Natural Science Foundation of Fujian Province (2024J102005), the Youth Talent Support Program of Fujian Province (Eyas Plan of Fujian Province) [2022], NSFC (62175205), NSAF (U2130112), and Shenzhen Science and Technology Development Funds (JCYJ20220530143015035). Z.D. would like to acknowledge the funding support from the Agency for Science, Technology and Research (A*STAR) under its MTC IRG (Project No. M22K2c0088), SUTD Kickstarter Initiative (SKI) grant with the award No. SKI 2021_06_05, and National Research Foundation via Grant No. NRF-CRP30-2023-0003. W. C. and J. S. would like to acknowledge the funding supporting from the China Scholarship Council Scholarship (CSC NO. 202306310153). and (CSC NO. 202306310168).

**Author contributions:**

C.W., G.Y., and Y.M. contributed equally to this work. Conceptualization: J.F., C.W. and G.Y. Methodology: C.W. and Y.M., J.Q. Fabrication: C.W. and J.Q. Characterization: J.F. and Z.G. Simulation: C.W. and Y.X. Discussions and suggestions: M.Y., J.F. and Z.G. Supervision: Z.G. and J.F. Writing—original draft: C.W., Z.G., and J.F. Writing—review and editing: Z.D. and J.F. All authors analyzed the data and read and corrected the manuscript before the submission.

All the authors discussed the results and commended to the manuscript. WC, SZ and CW contributed equally to this work.

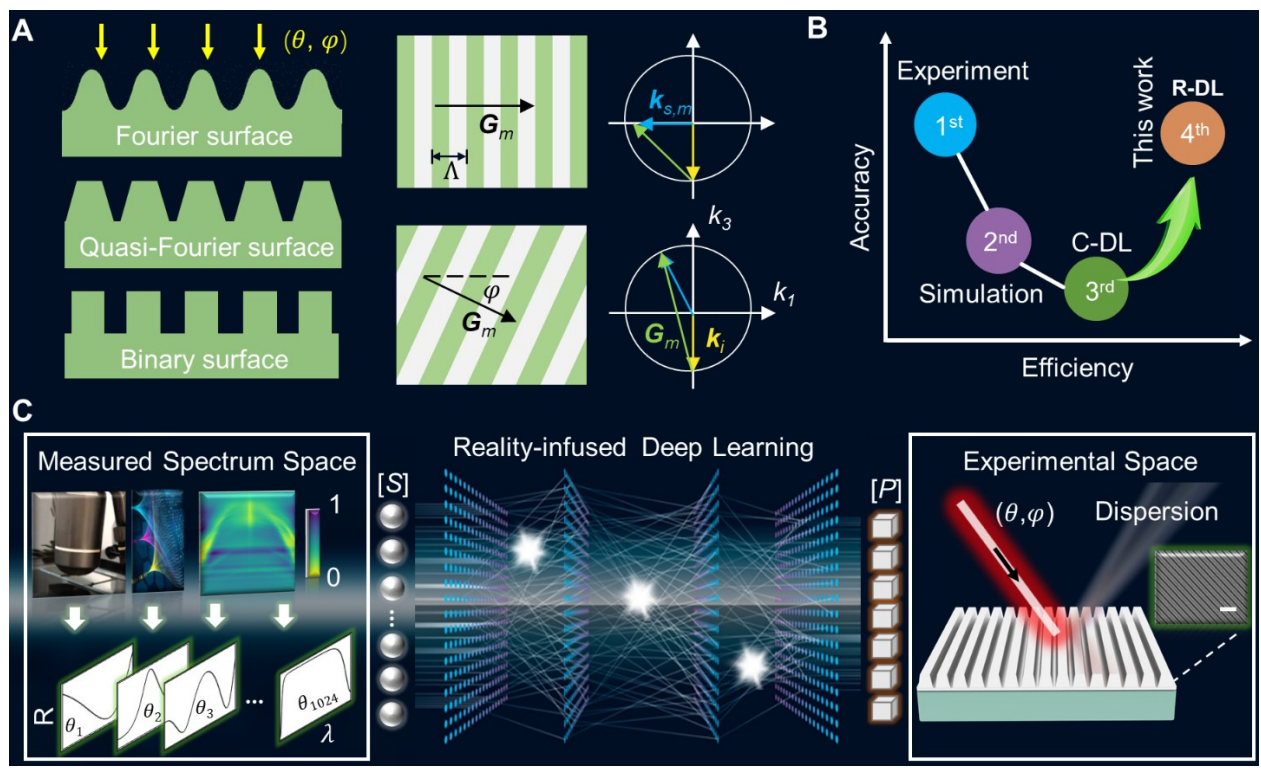

**Fig. 1. Transformation from simulation-driven to reality-infused deep learning. A** Left panel: Schematic diagram of optical Fourier surface, quasi-Fourier surface, and binary surface. Right panel: Desired waves in free space for metasurfaces at different incidence angles, and $k$-Vector diagrams of Horizontal and vertical models, where the symbols $k_i$, $G_m$, $k_{s,m}$ and $m$ represent the light vector, grating vector, $k$-vector of constructively scattered light, and diffraction order, respectively. **B** Evolution for various metasurface-related developing paradigms, C-DL denotes conventional deep learning and R-DL denotes reality-infused deep learning. **C** Schematic drawing of the R-DL architecture for the end-to-end design.

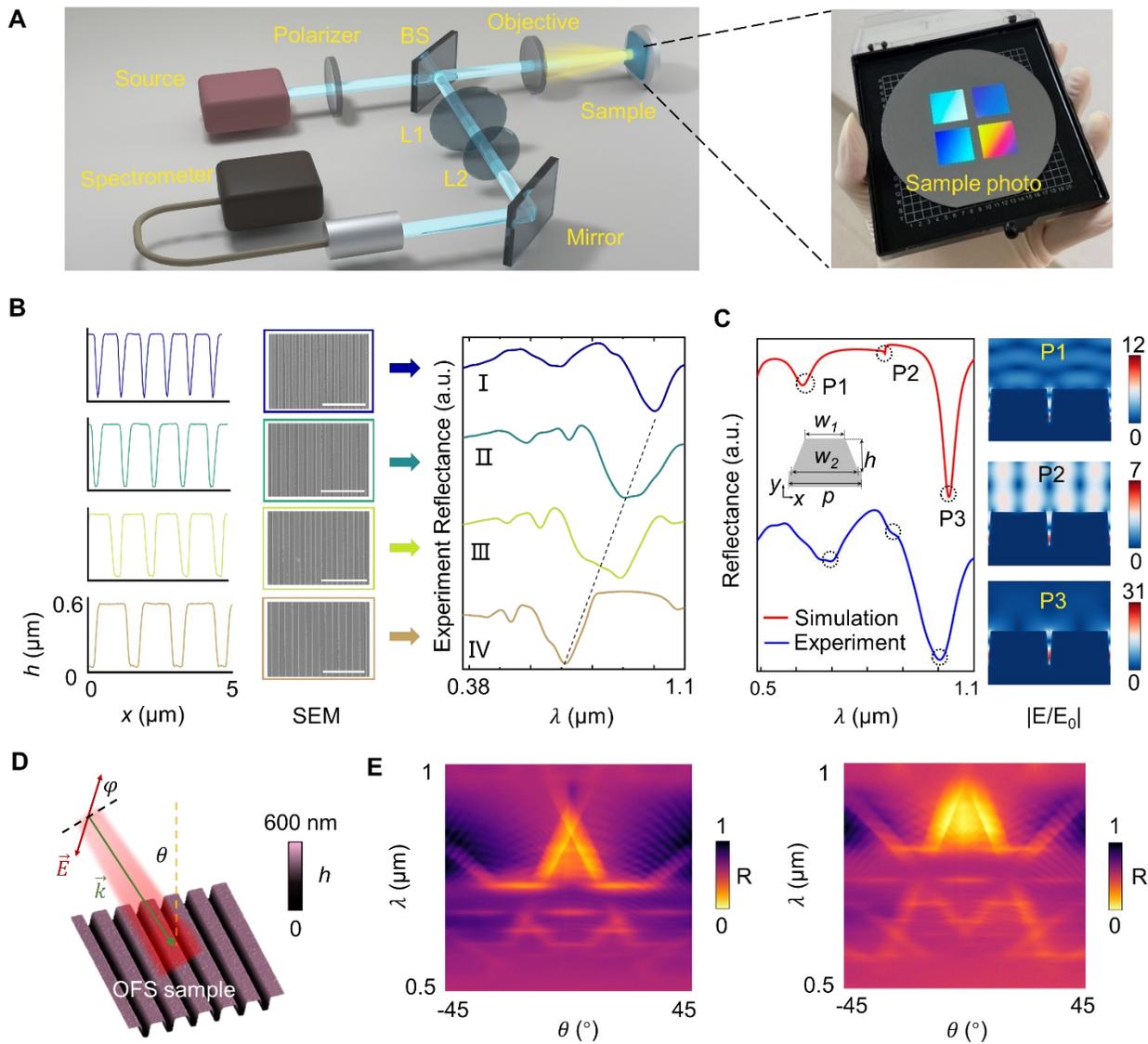

**Fig. 2. Measurement and simulation for angle-resolved quasi-OFSs. A** Schematic drawing of the ARS measurement and photograph of the wafer with four representative quasi-OFSs. **B** 2D AFM profiles, SEM images, and measured reflectance spectra for the four different samples (the scale bar is 5 μm). **C** Measured, simulated spectra, and electric field distributions of sample I with $w_1$=490 nm, $w_2$=740 nm, $p$=820 nm, $h$=575 nm. **D** 3D AFM image of the quasi-OFS sample and schematic of different incident wave vectors. **E** Measured incident angle dispersions of sample III and sample IV with $\varphi$=0°.

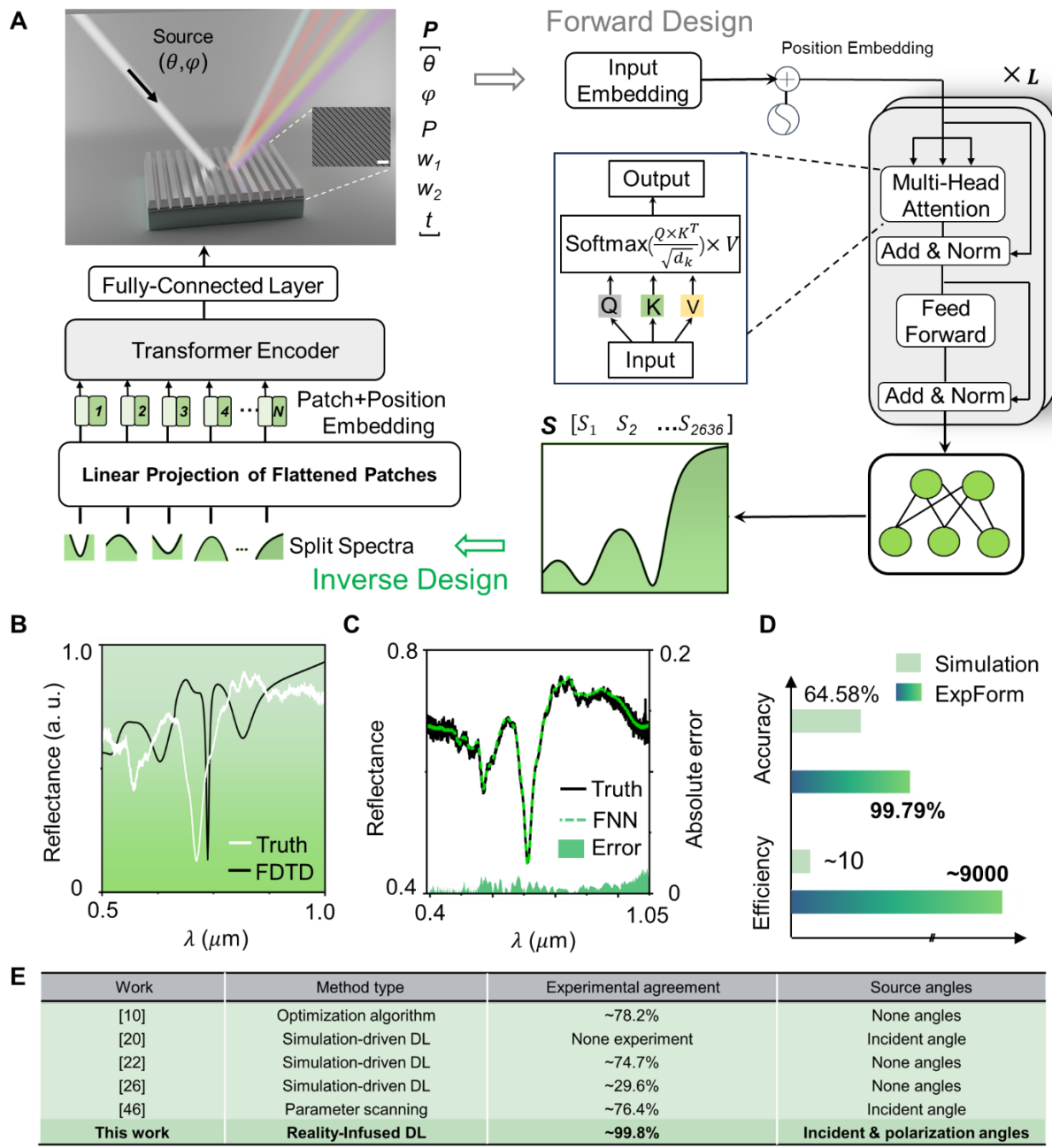

**Fig. 3. Performance comparison between simulation-based and reality-infused methods. A** Schematic of our ExpForm consisting with the FNN and INN. **B** Comparison of the measured and simulated spectra from one randomly selected instance. **C** Comparison of the measured and predicted spectra via our model. **D** Advantages of the R-DL scheme versus conventional simulation. **E** Comparison of state-of-the-art methods for optical design.

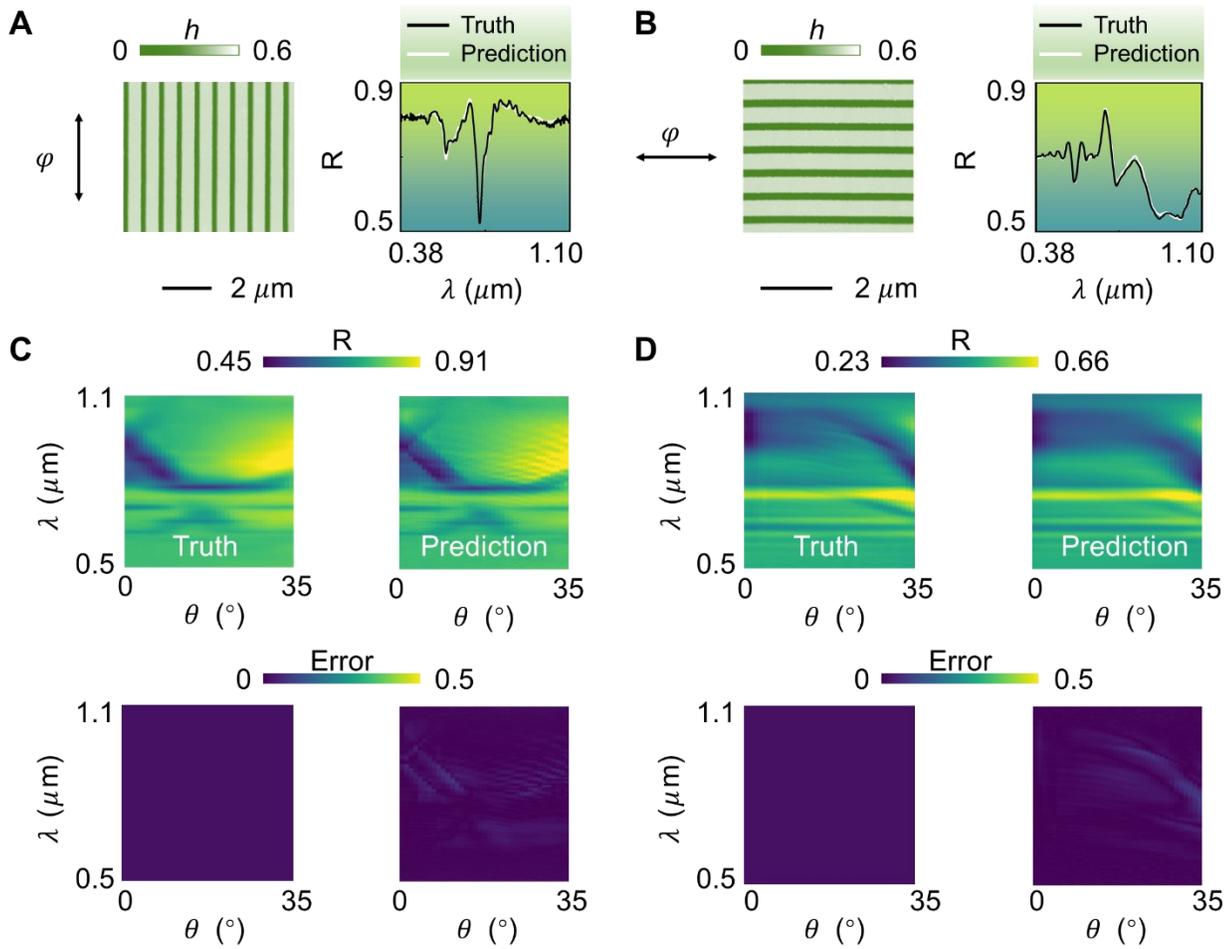

**Fig. 4. Angle-dispersion prediction for quasi-OFSs via our Expform. A, B** AFM diagrams and randomly-selected-angle spectra of two different samples. **C, D** Measured, predicted angle-resolved spectra and error distributions.

**Fig. 5. Inverse design flow of reality-infused deep learning.** On-demand design for a single resonance at wavelengths of **A** 700 nm and **B** 1000 nm. On-demand design for **C** high-reflection spectral profiles **D** and dual-band resonances. The illustrations are atomic force microscopy images.